\documentclass[3p,times,procedia]{elsarticle}
\usepackage{nupha_ecrc}

\volume{00}

\firstpage{1}

\journalname{Nuclear Physics A}

\runauth{L. Bianchi}

\jid{nupha}

\jnltitlelogo{Nuclear Physics A}

\usepackage{amssymb}

\usepackage{lineno}

\usepackage[figuresright]{rotating}

\newcommand{\Ks}{$K^{0}_{S}$}
\newcommand{\La}{$\Lambda$}
\newcommand{\aL}{$\overline{\Lambda}$}

\newcommand{\Xis}{$\Xi^{\pm}$}
\newcommand{\Oms}{$\Omega^{\pm}$}
\newcommand{\allparts}{\Ks, \La, \aL, \Xis~and~\Oms~}

\newcommand{\pt}{$\it{p}_{T}$}
\newcommand{\avpt}{$\langle\it{p}_{T}\rangle$}
\newcommand{\dnch}{${\langle dN_{ch}/d\eta \rangle}_{|\eta|<0.5}$}
\newcommand{\stev}{$\rm{\sqrt{s}~=~7~TeV}$}

\begin{document}

\begin{frontmatter}



\dochead{}

\title{Strangeness production as a function of charged particle multiplicity in proton-proton collisions}


\author{Livio Bianchi for the ALICE Collaboration}

\address{University of Houston, USA}

\begin{abstract}
Recent measurements performed in high-multiplicity proton-proton (pp) and proton-lead (p-Pb) collisions have shown features that are reminiscent 
of those observed in lead-lead (Pb-Pb) collisions. These observations warrant a comprehensive measurement of the production of identified particles.
We report on the production of \allparts at 
mid-rapidity measured as a function of multiplicity in pp collisions at $\sqrt{s}$ = 7 TeV with the ALICE experiment. Spectral shapes studied both 
for individual particles and via particle ratios such as ($\Lambda/K^{0}_{S}$) as a function of $p_{T}$ exhibit an evolution with event multiplicity 
and the production rates of hyperons are observed to increase more strongly than those of non-strange hadrons. These phenomena are qualitatively 
similar to the ones observed in p-Pb and Pb-Pb collisions.
\end{abstract}


\end{frontmatter}


\section{Introduction}
\label{sec:intro}

Strangeness production has been extensively studied in relativistic heavy-ion collisions as it represents an important tool to investigate
the properties of the strongly interacting system created in the collision, the so-called Quark-Gluon-Plasma (QGP).
Rafelski and M\"uller's expectation, first proposed in 1982 \cite{Rafelski_1982}, is that QGP formation should lead to a higher abundance of strangeness 
per participating nucleon than what is seen in pp interactions. 
Strangeness enhancement was indeed observed for the first time by comparing central heavy-ion collisions with proton-ion and pp reactions at 
the SPS \cite{WA97_1999}.\\
Heavy-ion experiments at RHIC and LHC have been exploited to investigate strangeness production at higher collision energies, revealing several 
features, such as the baryon/meson anomaly and lower enhancement at higher energy, which have been triggering further theoretical investigations (for a complete review 
see \cite{BLUME}). Quark coalescence has been used to explain the intermediate-\pt~baryon/meson enhancement \cite{GRECO}, while statistical
hadronization models succeeded in describing hadron yields 
\cite{BraunMunzinger200141} using a grand-canonical approach.
In the statistical model, strangeness production in small systems is affected by ``canonical suppression'', resulting in decreased
production rates compared to those measured in larger systems.\\
Recent observations by the ALICE experiment \cite{DIDIER,ALICE_Lk0spA} show that strangeness enhancement is also present in high-multiplicity p-Pb 
events and that the \La/\Ks~ratio shows a qualitatively similar behaviour in p-Pb and Pb-Pb collisions. Moreover, high multiplicity pp collisions at 
the LHC show features which, historically, were considered as signatures of QGP formation \cite{CMS_corrpp,ATLAS_ridge13}.
With the aim of investigating if QGP can be formed in high multiplicity pp collisions, the study of strangeness production is of primary importance.

\section{Analysis details and results} \label{sec:anadet}

The ALICE experiment \cite{ALICE_JINST,ALICE_PERF} detects \allparts through the reconstruction of the daughter tracks coming from their weak
decay in the rapidity region $|y|<0.5$. The Inner Tracking System (ITS) and the Time Projection Chamber (TPC) are used for primary and secondary 
vertex determination, tracking and particle identification through characteristic energy loss (dE/dx) measurements.\\
Low-pileup pp collisions at \stev~delivered by the LHC during 2010 have been used for this analysis. The set of $\sim$90M events was collected using a minimum-bias trigger, 
requiring a hit in either the V0 scintillators or in the SPD detector in coincidence with signals from beam pick-up counters. 
Events containing more than one primary vertex are tagged as pileup and are discarded.
Particle yields are measured applying a bin-counting technique after the application of several topological selections in order to decrease the combinatorial 
background. Acceptance and efficiency is determined through a PYTHIA 6 Perugia 0 Monte Carlo simulation with particle transport performed via a 
GEANT3 simulation of the ALICE detector. More details concerning the analysis procedure and systematics evaluation can be found in \cite{ALICE_strange09}.\\
The event multiplicity selection is performed using information from the V0 detector, which is composed by two arrays of scintillators placed at forward and 
backward rapidities. Multiplicity event classes are determined dividing the V0 amplitude distribution 
in percentiles and, for each class, the average pseudorapidity density of primary charged particles \dnch~is measured, 
as described in \cite{ALICE_pseudodenspA}. The fraction of the total systematic uncertainty which is uncorrelated 
over multiplicity is also estimated.

\begin{figure}[htb]
 \centering
 \includegraphics[scale=0.3062]{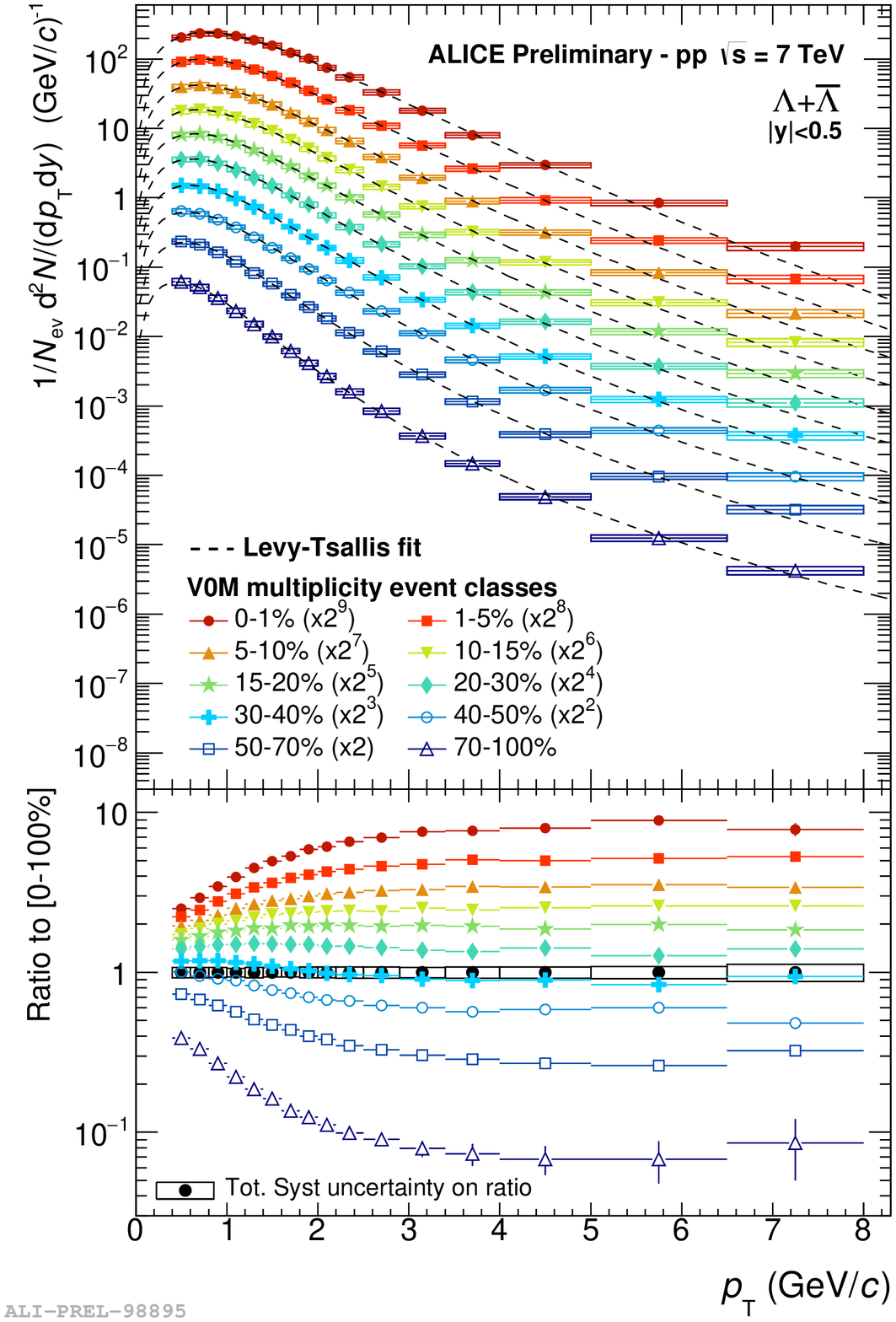}\hspace{1cm}
 \includegraphics[scale=0.3193]{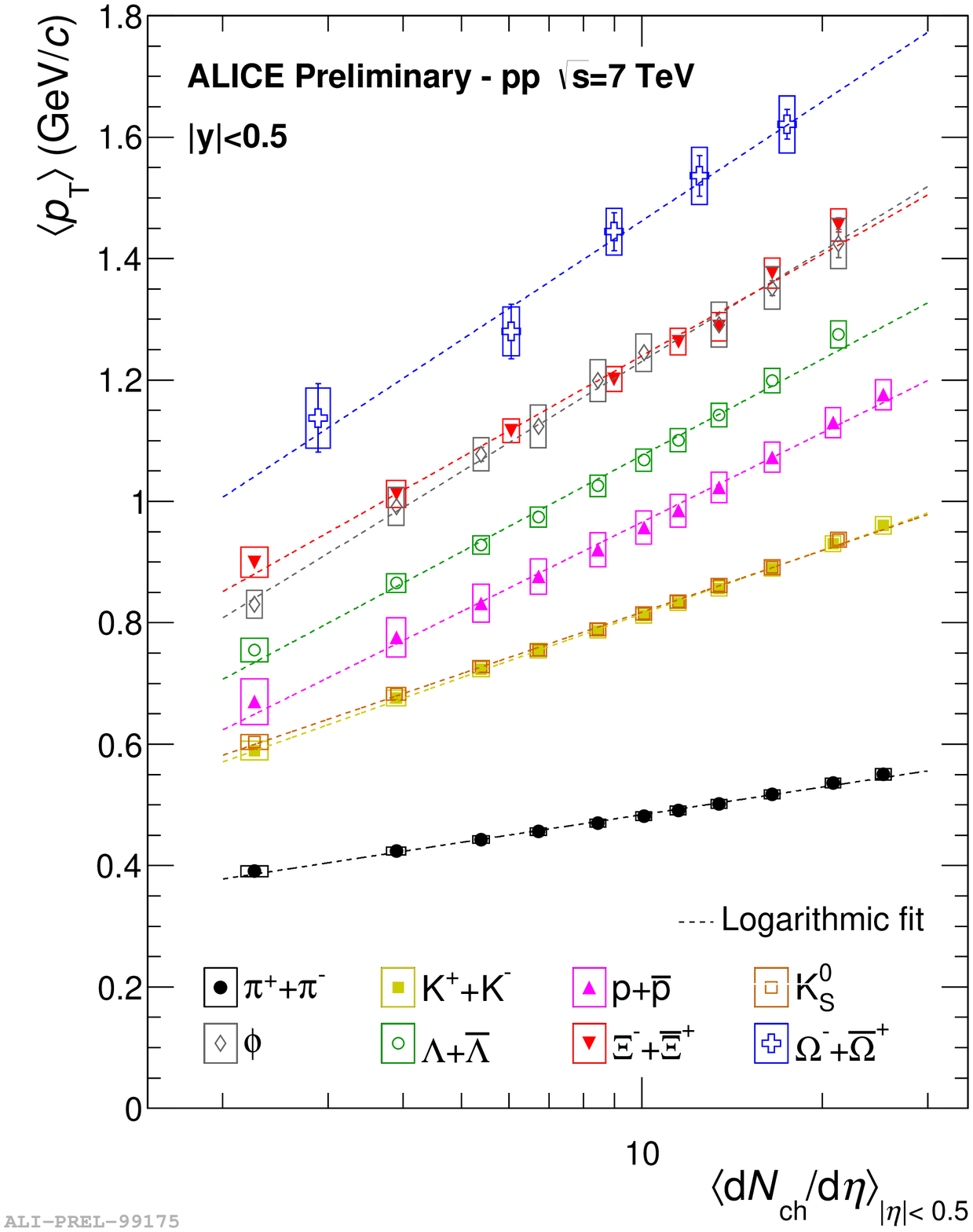}
 \caption{(left) \La+\aL~\pt~spectrum for different multiplicity bins and ratios to the 0-100\% spectrum. Boxes are total systematic uncertainties.
          (right) \avpt~as a funtion of event multiplicity for identified particles.}
 \label{fig:SpectralShape}
\end{figure}

\begin{figure}[htb]
 \centering
 \includegraphics[scale=0.45]{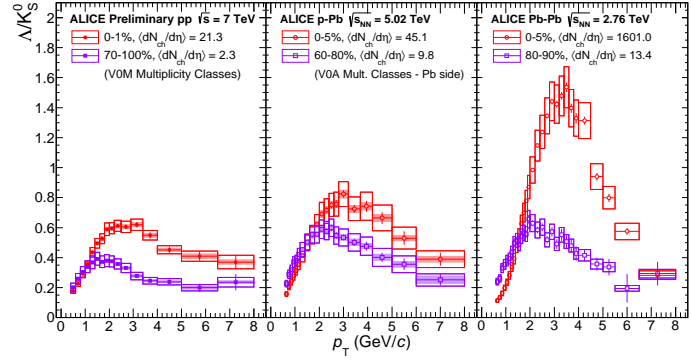}
 \caption{(\La+\aL)/\Ks~as a function of \pt~in the highest and lowest multiplicity classes considered in pp (left), p-Pb (centre) and
         Pb-Pb (right) collisions. Open boxes are total systematic uncertainties, shaded boxes show the uncorrelated-over-multiplicity fraction.}
 \label{fig:LambdaOverK0s}
\end{figure}

The measured transverse-momentum (\pt) distribution for \La+\aL~is shown in Fig.\ref{fig:SpectralShape}(left) for all the multiplicity classes 
considered. A hardening of the \pt~spectrum for increasing multiplicity is observed and the same effect is present for all the other particles 
considered. The evolution with multiplicity is concentrated at low transverse momentum, while at higher \pt~the ratio of the multiplicity-differential 
spectrum to the multiplicity-integrated one stays constant.
In order to quantify this hardening effect, the \avpt~as a function of \dnch~is shown in Fig.\ref{fig:SpectralShape}(right) for all identified 
particles studied by the ALICE experiment. The first moment of the distribution increases logarithmically with the multiplicity.\\
The \pt-differential \La/\Ks~ratio is shown in Fig.\ref{fig:LambdaOverK0s} in the lowest and highest multiplicity classes for pp (left), 
p-Pb (centre) and Pb-Pb (right) collisions. The ratio is qualitatively similar in the three collision systems. It is worth noting that highest 
and lowest multiplicity classes correspond to very different \dnch~in the different systems. In particular, the magnitude of the baryon/meson 
enhancement is qualitatively comparable in the 0-1\% pp, 60-80\% p-Pb and 80-90\% Pb-Pb multiplicity classes, where \dnch~is comparable 
($10\lesssim$\dnch$\lesssim20$).\\

\begin{figure}[htb]
 \centering
 \includegraphics[scale=0.36]{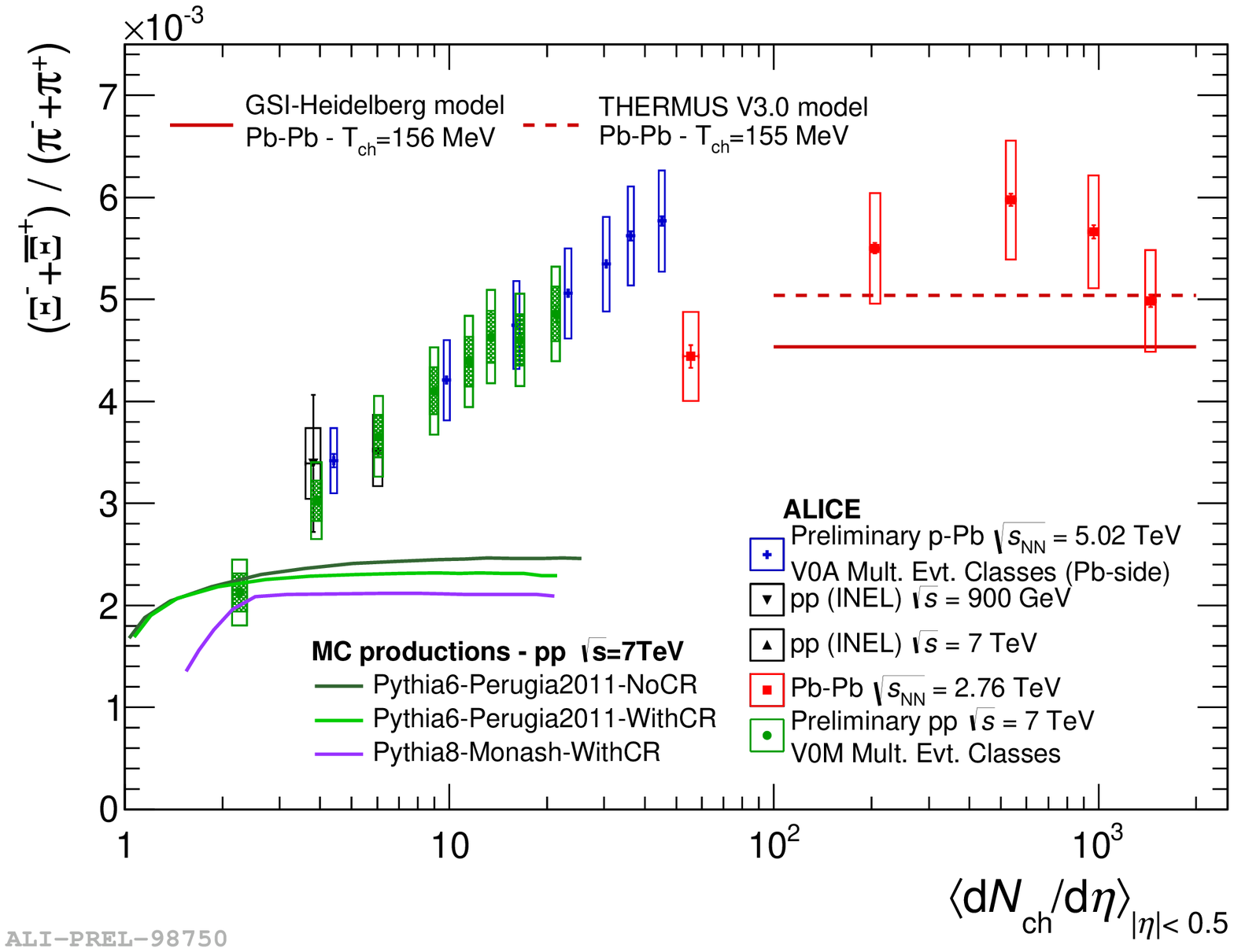}\hspace{0.5cm}
 \includegraphics[scale=0.36]{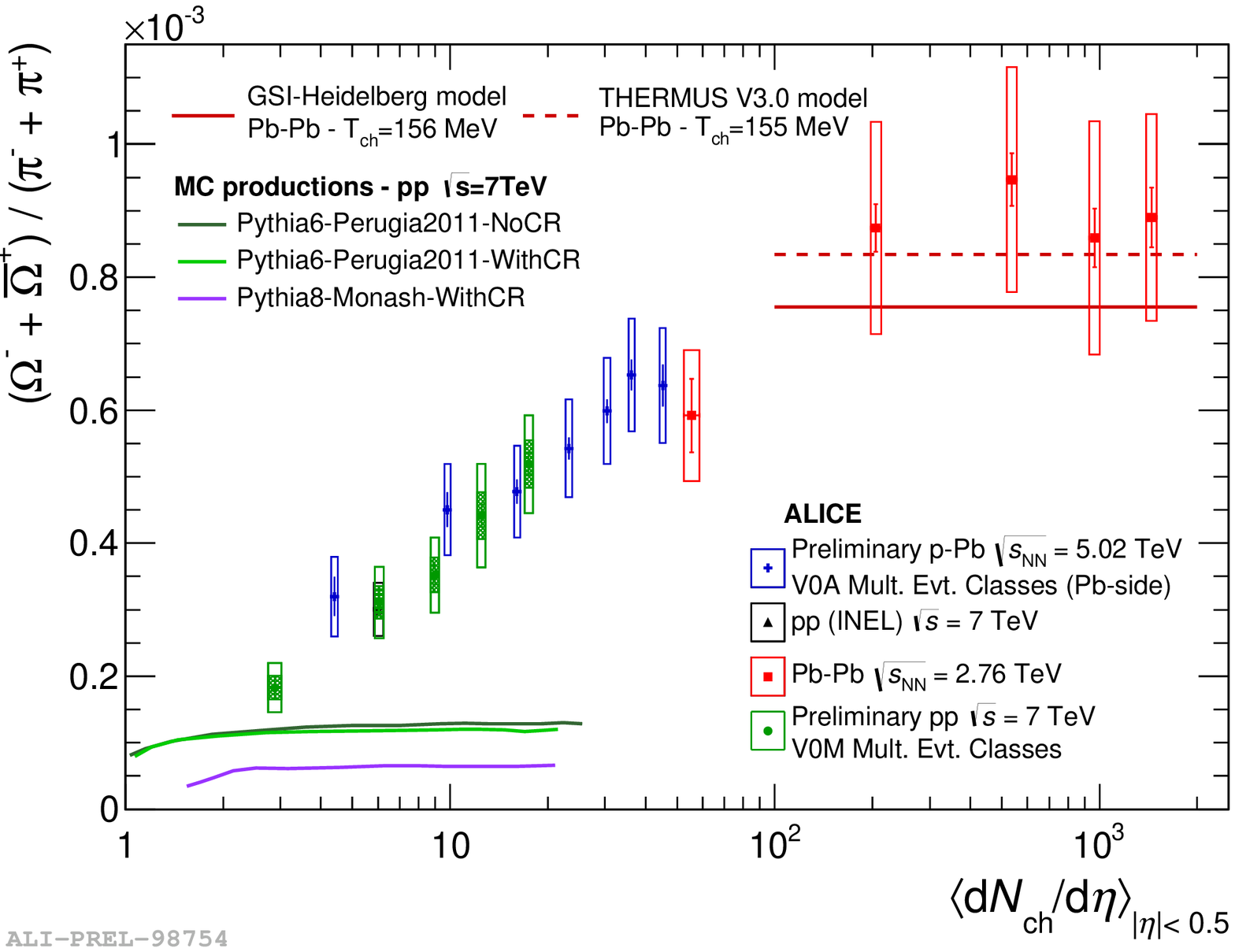}
 \caption{\Xis/$\pi^{\pm}$ (left) and \Oms/$\pi^{\pm}$ (right) ratios as a function of \dnch~for pp (green), p-Pb (blue) and Pb-Pb (red) collisions. 
          Open boxes are total systematic uncertainties, shaded boxes show the uncorrelated-over-multiplicity fraction.}
 \label{fig:FinalPlot}
\end{figure}

Particle yields are obtained integrating the \pt~distribution and extrapolating in the unmeasured low-\pt~region by means of a Levy-Tsallis fit to the 
full spectrum. \Ks~are measured down to zero \pt, while for \La+\aL, \Xis~and \Oms~the extrapolation fraction corresponds
respectively to 8(9)\%, 16(33)\%, 25(45)\% of the total yield in the lowest(highest) multiplicity bin. Blast-wave and Boltzman parametrizations are 
used to estimate the systematic uncertainty on the extrapolation.\\
The ratio of \Xis(\Oms) to $\pi^{\pm}$ yields is shown in Fig.\ref{fig:FinalPlot}left(right) as a function of the event activity as green
points for pp collisions. We observe a rising trend which is identical to the one observed in p-Pb collisions. In the case of 
\Xis~produced in high multiplicity events, the grand-canonical saturation limit calculated with two different implementations of the statistical model 
(THERMUSV3.0 \cite{Cley_2006} and GSI-Heidelberg model \cite{Andronic_2009}) is reached, while this is not the case for \Oms. 
The results are compared to MC calculations performed using three different versions of
the PYTHIA generator. None of the versions reproduce the observed trend and the inclusion of the color-reconnection mechanism doesn't introduce
any substantial improvement.
Same observation and conclusions hold for \Ks~and \La+\aL.

\begin{figure}[htb]
 \centering
 \includegraphics[scale=0.325]{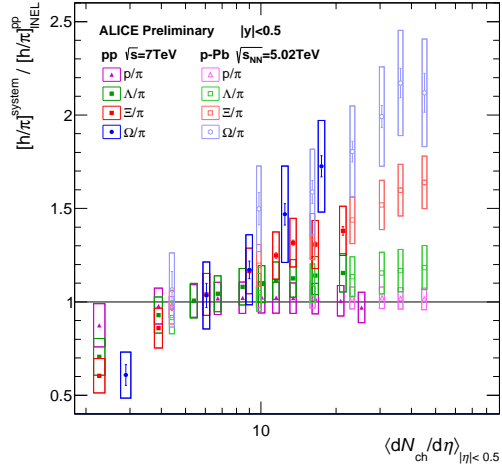}
 \caption{Particle yield ratios to pions for protons (p) and strange baryons normalized to the values measured in pp 0-100\%.}
 \label{fig:ScalingPlot}
\end{figure}

In Fig.\ref{fig:ScalingPlot} the yield ratios to pions normalized to the values measured for the pp inclusive sample is shown for all the measured strange 
particles and for protons. Note that here systematic errors are total, and are strongly correlated across multiplicity.
The relative increase with multiplicity is seen to be particle-dependent and, in particular, is more pronunced for baryons with higher strangeness content.
Proton yields do not increase more than pions when the multiplicity increases.

\section{Conclusions}
First results on strangeness production as a function of the charged particle multiplicity in \stev~pp collisions have been presented by the ALICE 
Collaboration. The \pt~distribution shows a clear evolution towards harder spectra at higher multiplicity for all the measured particles. 
The hardening saturates at $\sim$4 GeV/$\it{c}$. The \La+\aL/\Ks~ratio shows a qualitatively similar dependence on multiplicity for pp, p-Pb and 
Pb-Pb collisions when multiplicity classes corresponding to similar \dnch~values are compared.
Strange-baryon/pion yield ratios show an increase with multiplicity which is remarkably similar to the one observed in p-Pb collisions, suggesting a 
similar mechanism at play in the two systems. This increase is a strangeness-related effect since the proton/pion ratio is unaltered
with multiplicity and the relative increase is more important for baryons with higher strangeness content. These results cannot be reproduced by 
any of the commonly used PYTHIA tunes and are consistent with the hypothesis of QGP formation in high multiplicity pp events.\\





\bibliographystyle{elsarticle-num}
\bibliography{Biblio.bib}







\end{document}